\documentclass[12pt,a4paper]{article}
\usepackage{amsmath}
\usepackage{graphicx}
\begin{document}
\begin{titlepage}
\title{ Effects of the reflective scattering  in hadron production   at high energies}
\author{S.M. Troshin, N.E. Tyurin\\[1ex]
\small  \it Institute for High Energy Physics,\\
\small  \it Protvino, Moscow Region, 142281, Russia}
\normalsize
\date{}
\maketitle

\begin{abstract}
A gradual transition to the reflecting scattering mode developing already at the LHC energies is affecting multiparticle production dynamics, in particular, relation 
of the centrality with the  impact parameter values of $pp$--collisions. We discuss the  issues in the framework of the geometrical picture 
for the multiparticle production processes proposed by Chou and Yang. We  consider effects of  reflective scattering mode presence for the inclusive
cross-sections. 

\end{abstract}
\end{titlepage}
\setcounter{page}{2}

\section*{Introduction.}
Various aspects of the multiparticle production processes are nowadays under studies  at the LHC (cf. e.g. \cite{alice,atlas,cmsmult}). 
Those  being aimed 
to the searches of the  solution of the fundamental problems of QCD and  obtaining  the 
observables which could help to understand the mechanisms 
of confinement, formation of the quark-gluon plasma and properties of the nonperturbative QCD. 
There is a common opinion that these observables are related to and reflect properties of the deconfined transient state existing in the hadron and 
nucleus collisions for a very short transient time.  
A number of distinctive features  of multiparticle production processes can be understood using the most simple geometrical picture of
hadron interactions which is based on the properties of collisions in the impact parameter space. The geometrical approaches were proposed  by many authors,
for the most comprehensive considerations one can be referred to the Chou and Yang paper \cite{chya}. 

In this note we discuss how the geometry and characteristic features of the 
multiparticle production processes would be affected by the gradual transition to the reflective scattering mode. 
Such transition is expected
to start at the LHC energies.

The elastic scattering amplitude in the impact paramer representation can cross   the ``black disk limit'' 
at small impact parameters and very high energies. Such possibility 
was considered many years ago in the paper  \cite{bbdl} where transition to the ``black ring picture'' 
has been anticipated  in the general framework of the rational form of unitarization on the base of  the CDF data 
obtained at  Tevatron. The value of the energy
where the amplitude reaches ``black disk limit'' was estimated later in \cite{laslo} and the approximate correspondence
between the two estimations performed in   \cite{bbdl} and \cite{laslo} has been established.
The energy dependence of the amplitude resulting in the crossing of the black disk limitation is a manifestation
of a gradual transition to the reflective scattering mode introduced and discussed in \cite{intje}.
Recent measurements of the differential cross-section of the elastic $pp$--scattering at the LHC \cite{totem}
have led to reopening (cf. e.g. \cite{dremin,anisovich}) of the theoretical discussion
on the above mentioned regime in elastic scattering and the recent straightforward and model-independent
reconstruction of the impact-parameter 
dependent elastic amplitude has clearly indicated that transition to
the reflective scattering mode has already  taken place  at the LHC energy of 7 TeV \cite{alkin}. 

It has been shown 
that the proton interaction region evolved
with growing energy from
the BEL (Blacker, Edgier, Larger)  \cite{henzi} picture at lower energies to the REL (Reflective, Edgier, Larger) picture at the
LHC energy of 7 TeV.
The appearance of the latter mode can be explained by the fact that opening of the
new inelastic channels with growing energy, does not lead
to saturation of the total probability of the inelastic collisions at small transverse distance, 
but instead,  results in the destructive interference and  
the self-damping of the inelastic channels \cite{intje,bakbl}.  
The natural question  is how the LHC data
can be affected, namely, what are direct or indirect results of this reflective scattering mode at the LHC energies and beyond. 

We consider here the implications of the above mentioned issues for the observables related to  many--particle 
production dynamics, in particular, multiplicity distribution of the secondary particles and centrality in $pp$-collisions.

\section{Reflective scattering mode and multiparticle production.}
The most essential feature of the reflective scattering mode is the peripheral impact profile of the inelastic overlap function 
\begin{equation}\label{hin}
 h_{inel}(s,b)\equiv \frac{1}{4\pi}\frac{d\sigma_{inel}}{db^2} 
\end{equation}
which enters the unitarity equation for the elastic scattering
amplitude $f(s,b)$:
\begin{equation}\label{unit}
  \mbox{Im} f(s,b)=h_{el}(s,b)+h_{inel}(s,b).
\end{equation}

The function $S(s,b)=1+2if(s,b)$ is the $2\to 2$ 
elastic scattering matrix element. 
Since the following discussion is the most qualitative one, we take  the   amplitude $f(s,b)$ to be a pure imaginary function, replacing 
$f\to if$. The function $S(s,b)$  is a real one, but can  change its sign. 
The maximum value of $h_{inel}(s,b)=1/4$
can be reached at high energies at the positive non-zero values of the impact parameter at $b=R(s)$.
The numerical analysis \cite{alkin} clearly indicates that  it happens at $R(s)\simeq 0.2$ $fm$ at the energy $\sqrt{s}=7$ $TeV$.
Energy dependence of the function $R(s)$ is usually described by the logarithmic function of energy,  i.e. $R(s)\sim\frac{1}{\mu} \ln s$.
Such a dependence takes place in the most phenomenological
models. It is consistent with the analytical properties of the elastic scattering amplitude and mass $\mu$ 
might be related to the mass of pion.
Since the derivatives of $h_{inel}(s,b)$ have the forms:
\begin{equation}\label{der}
 \frac{\partial h_{inel}(s,b)}{\partial s}=S(s,b)\frac{\partial f(s,b)}{\partial s}\,\,\, \mbox{and}\,\,\,
 \frac{\partial h_{inel}(s,b)}{\partial b}=S(s,b)\frac{\partial f(s,b)}{\partial b},
\end{equation}
it is evident that 
\begin{equation}\label{max}
 \frac{\partial h_{inel}(s,b)}{\partial b}=0
\end{equation}
at $b=R(s)$. $S(s,b)=0$ at $b=R(s)$ by the definition of the function  $R(s)$, i.e. the complete absorption of 
the initilal elastic channel takes place at the value of the impact parameter $b=R(s)$ . 
\begin{figure}[h]
\begin{center}
\resizebox{8cm}{!}{\includegraphics*{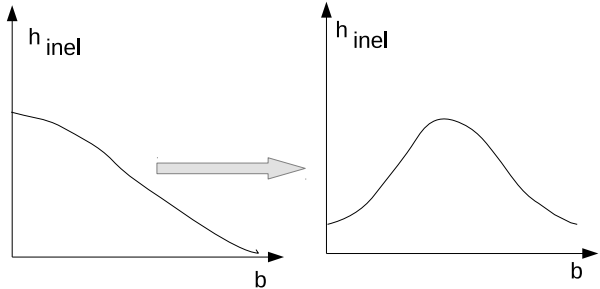}}
\end{center}
\caption[ch1]{\small Transition of the inelastic overlap function with growing energy from a central to a  peripheral profile.}
\end{figure}
This transition from a central to a peripheral form would be slowing the energy dependence of the observables related to 
the multiparticle production processes. In particular, slow down of the mean multiplicity is expected at the highest values
of the LHC energies \cite{slow}.
Evidently, the derivative of the inelastic overlap function has  the sign opposite to the sign 
of $\partial f(s,b)/\partial b$ in the region where $S(s,b)<0$. It is  the region of the $s$ and $b$ variables
where the function $S(s,b)$ is negative
(the  phase of $S(s,b)$ is such that $\cos 2\delta(s,b)=-1$) is
responsible for the transformation of the central impact profile of the function $f(s,b)$ into a peripheral profile 
of the inelastic overlap function $h_{inel}(s,b)$ (Fig.1).  
Thus, at the sufficiently  high energies ($s>s_0$),
the two separate  regions of
 impact parameter distances can be anticipated, i.e. the outer region
of peripheral collisions with scattering of a typical absorptive origin, i.e.
$S(s,b)|_{b>R(s)}>0$ and
 the inner region of central collisions
where the scattering has a combined reflective and absorptive nature, $S(s,b)|_{b< R(s)}<0$.
 The function $S(s,b)$ can be rewritten in the form
\begin{equation}\label{ssb}
 S(s,b)=\kappa(s,b)\exp[2i\delta(s,b)],
\end{equation}
where $\kappa(s,b)$ and $\delta(s,b)$ are the real functions. The function $\kappa(s,b)$  
($0\leq \kappa(s,b) \leq 1$)
 is called an absorption factor which is
related to the contribution of the inelastic channels into unitarity relation, its zero value, $\kappa=0$, corresponds to a complete 
absorption of the incoming channel. The interpretation of this factor, in fact, depends on the value of the phase $\delta(s,b)$.
The transition to the negative
values of $S$ leads to
appearance of the real   phase shift, i.e. $\delta(s,b)|_{b< R(s)}=
\pi/2$ \cite{bbdl}.
The value of this factor is determined by the inelastic channels contribution
to the unitarity equation for the elastic scattering amplitude $f(s,b)$, i.e.
\begin{equation}\label{kap}
 \kappa^2(s,b)=1-4h_{inel}(s,b),
\end{equation}
It can be easily seen by expressing the function $h_{inel}(s,b)$ as a product, i.e
\begin{equation}\label{inel}
h_{inel}(s,b)=f(s,b)[1-f(s,b)]. 
\end{equation}
When $f(s,b)\to1$ the inelastic overlap function $h_{inel}(s,b)$ becomes the most peripheral one while the elastic overlap function
$h_{el}(s,b)$ remains to be central. Therefore, elastic scattering at large values of $-t$ is dominated by the pure elastic
reflective scattering, while at the large impact parameters (note, that the amplitude $f(s,b)$ is small in this limit) the following approximate relation is valid
\begin{equation}\label{larb}
 f(s,b)\simeq h_{inel}(s,b).
\end{equation}
This relation means that the diffraction peak in elastic scattering at small values of $-t$  results from the multiparticle production
processes dynamics in this region. We have used here a  qualitative kinematical corresponence between small values of $-t$ and large values of $b$
and vice versa. It is not surprising therefore that elastic scattering at small and large values of $-t$ would have different dependencies
on the transferred momentum since the scattering processes in these two regions are determined by the different dynamical mechanisms, namely, 
one is determined by absorption  and another one --- by reflection.

Typical asymptotic patterns of  elastic and inelastic overlap functions 
are depicted on Fig. 2.
\begin{figure}[hbt]
\begin{center}
\resizebox{12cm}{!}{\includegraphics*{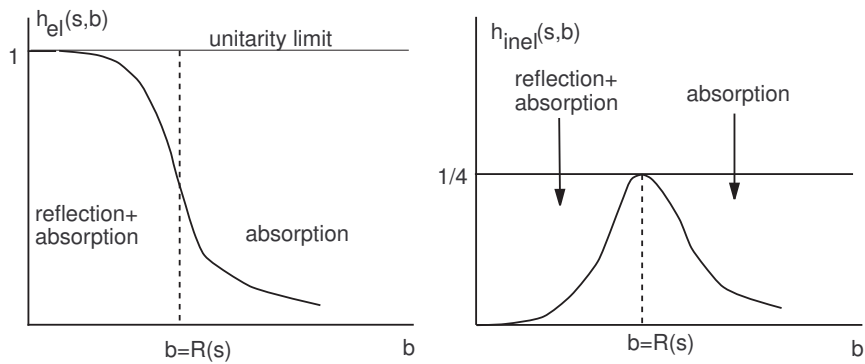}}
\end{center}
\caption[ch2]{\small Typical qualitative forms of the impact parameter profiles relevant for the
elastic and inelastic overlap functions at the
 asymptotically high energies.}
\end{figure}
It should be noted that the probability of an inelastic process in the hadron collision at the energy $s$ and impact parameter $b$ is the following
\begin{equation}\label{pinel}
  P_{inel}(s,b)=4h_{inel}(s,b)\equiv  { \frac{d\sigma_{inel}}{2\pi bdb}}
\end{equation}
and
\begin{equation}\label{inel1}
\sigma_{inel}(s)=2\pi\int_0^\infty  P_{inel}(s,b)bdb.
\end{equation}
Therefore, any observable, which desribe a multiparticle production process, $A(s,\xi)$ ($\xi$ is a variable or a set of  variables), 
can be obtained from the corresponding impact-parameter dependent function $A(s,b,\xi)$
by integrating it with the weight function  $h_{inel}(s,b)$, e.g.
\begin{equation}\label{mm}
 A (s,\xi)=\frac{\int_0^\infty A(s,b,\xi) h_{inel}(s,b)bdb}  
{\int_0^\infty h_{inel}(s,b)bdb} .
\end{equation}
Thus, on the grounds of the prominent peripheral dependence of $h_{inel}(s,b)$ (it has a peak at $b=R(s)$)
at asymptotically high energies (cf. Fig. 2)
one can obtain the following approximate relation
valid in the limit $s\to\infty$:
\begin{equation}\label{mst}
A (s,\xi)\simeq A(s,b,\xi)|_{b=R(s)}.
 \end{equation}
 This relation is applicable for many observables associated with particle production proceses  when the reflective scattering dominates
 at very high energies. In particular, it can be applicable to the mean multiplicity, average transverse momentum, anisotropic flows $v_n$ and multiplicity
 distribution $P_n(s)$. In general, it means that in the multiparticle production processes the relative range of the variations of the impact parameter 
 is decreasing with energy and the typical inelastic event at very high energy is the event with the non-zero value of the impact parameter in the region around $b=R(s)$ while the inelastic events at small and large impact parameter values are strongly suppressed.
 However, it is not the case for elastic collisions, where impact parameter profile being a central  one (Fig. 2).
 
 It should be noted that the average values of the impact parameters $\langle b^2\rangle(s)_{el}$ and $\langle b^2\rangle(s)_{inel}$ 
 for elastic and inelastic collisions have the similar asymptotic energy dependencies at $s\to \infty$, i.e.
 \begin{equation}\label{imp}
\langle b^2\rangle_{el}(s) , \,\, \langle b^2\rangle_{inel}(s)\sim R^2(s),
  \end{equation}
  while the ratios 
  \begin{equation}\label{rat}
     \sigma_{el}(s)/\langle b^2\rangle_{el}(s)\sim const.,\,\,\,\sigma_{inel}(s)/\langle b^2\rangle_{inel}(s)\sim 1/\ln(s)
  \end{equation}
and as it is evident, behave differently at $s\to\infty$. The second relation corresponds to decrease with energy 
of the relative range of the impact parameter fluctuations
in the inelastic processes, namely $\Delta b/\langle b \rangle$ decreases as $1/\ln s$ for such processes.

 We consider now the multiplicity distribution function $P_n(s)$,
 \begin{equation}\label{pn}
   P_n(s)\equiv \sigma_n(s)/\sigma_{inel}(s), 
 \end{equation}
where  $\sigma_n(s)$ is  the $n$-particle production cross-section . 
According to Eq. (\ref{mst})  $P_n(s)$ can be written in the form
\begin{equation}\label{md}
P_n(s)\simeq P_n(s,b)|_{b=R(s)}={ \frac{d\sigma_n}{2\pi bdb}}|_{b=R(s)}
 \end{equation}
since 
\[
 { \frac{d\sigma_{inel}}{2\pi bdb}}|_{b=R(s)}=1.
\]

\section{What is the centrality in $pp$--collisions?}
 In nuclear reactions, the results of the measurements are presented for various centralities.
 This variable is determined as a sum (cf. \cite{centrality} for the recent discussion of this quantity).
 \begin{equation}\label{cnt}
 c_N(s)=\sum_{n=N}^\infty P_n(s).
 \end{equation}
 If one can correlate centrality with the impact parameter, this variable can be used for description of the 
 collision geometry. The relation of centrality with impact parameter
 has been obtained for the nuclear reactions in \cite{bron} :
 \begin{equation}\label{central}
  c_N(s)\simeq \frac{\pi b^2(N)}{\sigma_{inel}(s)},
 \end{equation}
 and the same relation was considered to be valid in case of
 $pp$--interactions in \cite{dirf}.
 Here $b(N)$ is the impact parameter value where the mean multiplicity $\langle n \rangle (s,b)$ is equal to $N$.
 In both cases it was supposed that the black disk geometrical picture of the collisions 
 is valid. 
 
 This assumption ceases to be true when a gradual transition to the reflective scattering mode starts. As it was already
 mentioned, it is already experimentally observed   at $\sqrt{s}=7$ TeV \cite{alkin}. 
 To obtain an expression for centrality in case of the reflective scattering domination, one can apply Eq. (\ref{md}) for $P_n(s)$, i.e.
 \begin{equation}\label{cntr}
  c_N(s)\simeq \sum_{n=N}^\infty \frac{d\sigma_n}{2\pi bdb}|_{b=R(s)}=1- \sum_{n=3}^N\frac{d\sigma_n}{2\pi bdb}|_{b=R(s)},
 \end{equation}
since
\begin{equation}\label{inel2}
  \sum_{n=3}^\infty\frac{d\sigma_{n}}{2\pi bdb}= \frac{d\sigma_{inel}}{2\pi bdb}.
\end{equation}
Thus, when the reflective scattering dominates at $s\to\infty$, centrality $c_N$ is not a measure of the impact parameter, but it is
associated with the  value of  $b=R(s)$, where absorption is maximal. Centrality is then related to the dynamics
of the multiparticle production processes when impact parameter can variate in  the relatively narrow range around 
the value $b=R(s)$. 

To proceed further
and to get an additional information on centrality we need information on the impact parameter dependence of the cross--sections 
in multiparticle production and we
would like to apply for that purpose the geometrical picture of hadron production which has been 
mentioned in the Introduction. 
\section{Geometrical picture and multiparticle production.}
As  it is evident from Fig. 2, elastic scattering occurs in the wide range  of the impact parameter  variatons. The same is true and for the inelastic processes, but at lower energies only (cf. Fig. 1). 
It was shown \cite{chyang} that
a wide range of angular momentum produces a coherent superposition and results in the high forward elastic peak observed in the data from the low energies till the LHC ones.
In \cite{chya} it was proposed that this wide range of impact parameter variations is responsible for the nonstochastic
aspects of the multiparticle production dynamics leading to the KNO scaling \cite{kno} or negative-binomial distribution of the 
multiplicity distribution function. 

The following geometric mechanism has been proposed in \cite{chya}:
the broad multiplicity distribution in hadron production has been related to the result of the incoherent superposion of the collisions with the different
 impact parameter values. In contrast with above, collisions at fixed  impact parameter value  were supposed to lead a narrow multiplicity distribution. However, as it was shown in \cite{alkin} the reflective scattering
 mode has already been detected at $\sqrt{s}=7$ TeV. Starting from the indicated energy value this  mode is expected to become more and more prominent leading
 asymptotically to Eq. (\ref{mst}). The relative range of the impact parameter variations will be decreasing with the energy increase and gradual transition
 from the nonstochastic multiplicity distribution to the stochastic one should be expected.
 
 Combining reflective scattering with the mechanism of Chou and Yang \cite{chya}, one can conclude that asymptoticaly  the 
 multiplicity distribution would be a product of two Poisson--like distributions one in $n_F$ and another one in $n_B$ ($n_F$ and $n_B$-the multiplicities of the secondary 
 particles in forward and backward hemispheres) both with the average multiplicity $\langle n \rangle (s,b)|_{b=R(s)}$.
It should be also noted  that since particle production in forward and backward hemispheres   occurs at the same value of the impact parameter, it is predicted that $n_F \simeq n_B$ and the correlation parameter $b_c$ defined as
\begin{equation}\label{corr}
b_c=\frac{\langle n_Fn_B \rangle-\langle n_F \rangle\langle n_B \rangle}{\langle n_F^2 \rangle-\langle n_F \rangle^2}
 \end{equation}
tends to unity at $s\to \infty$. This prediction remains valid when reflective mode is present. 
It seems interesting to check this prediction at the LHC energies. 

It should be noted that the reflective sattering mode dominance would affect inclusive cross--section which has the following form (cf. \cite{dpe} for details and notions) in the rational form of unitarization:
\begin{equation}\label{incl}
E\frac{d\sigma}{d^3q}=8\pi\int_0^\infty bdb \frac{I(s,b,q)}{|1-iU(s,b)|^2} \equiv 8\pi\int_0^\infty bdb \frac{I(s,b,q)}{\mbox {Im}U(s,b)}h_{inel}(s,b)
 \end{equation}
and
\begin{equation}\label{sum}
 \int \frac{d^3q}{E}I(s,b,q)=\langle n\rangle (s,b)\mbox{Im} U(s,b).
\end{equation}
The reflective scattering appears
naturally
 in the $U$--matrix (rational) form of unitarization which has been used above for the inclusive cross--section.
In the $U$--matrix approach,
 the $2\to 2$ scattering matrix element in the
impact parameter representation
is the following linear fractional transform:
\begin{equation}
S(s,b)=\frac{1+iU(s,b)}{1-iU(s,b)}. \label{um}
\end{equation}
 $U(s,b)$ is the generalized reaction matrix, which is considered to be an
input dynamical quantity. This relation (\ref{um}) is one-to-one transform and  can  easily be
invertible. Such form of unitarization can result from from the cofinement condition for the 
colored degrees of freedom\cite{confinement}.
Inelastic overlap function $h_{inel}(s,b)$
can be expressed through the function $U(s,b)$ by the following relation
\begin{equation}\label{hiu}
h_{inel}(s,b)=\frac{\mbox{Im} U(s,b)}{|1-iU(s,b)|^{2}},
\end{equation}
and the only condition to obey unitarity by the elastic scattering amplitude 
 is $\mbox{Im} U(s,b)\geq 0$. Elastic overlap function is related to the function
 $U(s,b)$ as follows
\begin{equation}\label{heu}
h_{el}(s,b)=\frac{|U(s,b)|^{2}}{|1-iU(s,b)|^{2}}.
\end{equation}
The form of $U(s,b)$ depends on the particular model chosen,
for our qualitative
 purposes it
is taken to be an  increasing function with energy in a power-like way
 and decreases with impact parameter
like a linear exponent or Gaussian.
To simplify the qualitative picture, we consider also the function $U(s,b)$
as  a pure imaginary.

Again, starting from the prominent peripheral dependence of $h_{inel}(s,b)$ 
at asymptotically high energies, 
one can obtain the following approximate relation for the inclusive cross-section (we have supposed that $U(s,b)$ is pure imaginary function
and have taken into account that $U(s,b)|_{b=R(s)}=1$)
valid in the limit $s\to\infty$:
\begin{equation}\label{incl1}
E\frac{d\sigma}{d^3q}\simeq I(s,b,q)|_{b=R(s)}\sigma_{inel}(s),
 \end{equation}
where
\begin{equation}\label{multp}
 \int \frac{d^3q}{E}I(s,b,q)|_{b=R(s)}\simeq \langle n\rangle (s).
\end{equation}
Thus, to obtain inclusive cross--section at very high energies, one should perform modelling of inelastic hadron interactions at the  value
of the  impact parameter of the colliding hadrons  around $b=R(s)$, where the complete absorption occures. The impact parameter $\mathbf {b}$ 
is a weighted sum of the  impact parameters of the final particles
\[
 \mathbf {b}=\sum_{i=1}^n x_i \mathbf {b}_i,
\]
where $x_i$ is a Feynman $x$ of a final particle $i$.

To proceed further in this way of modelling we consider one particularly interesting class of the multiple production processes, i.e. the double--pomeron echange (dpe) processes of the type
\[
 pp\to p+X+p,
\]
where plus corresponds to a gap in rapidity and $X$ is a system (or a particle) of particles deprived of a rapidity gap.
According to Eq. (\ref{incl1}), the inclusive cross--section of the dpe--process can be written in the form
\begin{equation}\label{idpe}
E\frac{d\sigma _{_{dpe}}}{d^3q}\simeq I_{dpe}(s,b,q)|_{b=R(s)}\sigma_{inel}(s).
 \end{equation}
This cross--section corresponds to the selection from the set of all final states $|n\rangle$ of the states $|n\rangle _{dpe}$ relevant for the 
double-pomeron exchange processes. To construct the expression for the function $I_{dpe}(s,b,q)$ we use notions described in \cite{dpe} and write
it as a convolution of the two distributions of the condesates in the colliding  protons and condensate (or parton) cross-section:
\begin{equation}\label{conv}
I_{dpe}(s,b,q)=\int D_c( \mathbf {b}_1)\sigma _0(s,  \mathbf {b}- \mathbf {b}_1- \mathbf {b}_2,q)D_c( \mathbf {b}_2)d \mathbf {b}_1d \mathbf {b}_2,
 \end{equation}
where $\sigma _0(s,  \mathbf {b}-\mathbf {b}_1-\mathbf {b}_2,q)$ is the (point-like) cross--section of condensate interaction written in the form
\begin{equation}
 \sigma _0(s,  \mathbf {b}-\mathbf {b}_1- \mathbf {b}_2,q)=\tilde{\sigma} _0(s, q)\delta( \mathbf {b}-\mathbf {b}_1-\mathbf {b}_2).
\end{equation}
Thus, $I_{dpe}$  can be written as a convolution
\begin{equation}\label{conv1}
I_{dpe}(s,b,q)=\tilde{\sigma }_0(s,q)D_c\otimes D_c= \tilde{\sigma }_0(s, q)\int D_c( \mathbf {b}-\mathbf {b}_1)D_c( \mathbf {b}_1)d \mathbf {b}_1
\end{equation}
We can assume that the condensate distribution in the hadron is controlled by the pion mass and $I_{dpe}$ has therefore the form
\begin{equation}\label{pion}
 I_{dpe}(s,b,q)\simeq\tilde{\sigma }_0(s,q)e^{-m_\pi b}.
\end{equation}
Since $R(s)=\frac{1}{\mu}\ln s$, we will have a suppression factor decreasing with the energy
\begin{equation}\label{pion1}
 I_{dpe}(s,b,q)|_{b=R(s)}\simeq\tilde{\sigma} _0(s,q) s^{-m_\pi/\mu}.
\end{equation}
This factor arises due to  the reflective scattering mode dominating at $s\to\infty$ and enters to the expression for the $dpe$ inclusive cross--section affecting
its energy dependence given by a product of the functions $\tilde{\sigma} _0(s,q) $ and $ \sigma_{inel}(s)$, i.e.
\begin{equation}\label{idpe1}
E\frac{d\sigma _{_{dpe}}}{d^3q}\simeq \tilde{\sigma} _0(s,q) \sigma_{inel}(s) s^{-m_\pi/\mu},
 \end{equation}
where $ \sigma_{inel}(s)\sim \ln s $ at $s\to \infty$ when the reflective mode becomes dominating one. Thus, in this energy limit  the double--pomeron exchange 
processes would not survive unless $\tilde{\sigma} _0(s,q) $ is growing with energy at least as a power of it, i.e. $\tilde{\sigma} _0(s,q)\sim s^\alpha $ with
$\alpha \geq -m_\pi/\mu$ (we neglect here the logarithmic-type dependencies).
\section*{Conclusion.}
In general, the reflective scattering mode would change nonstochastic dynamics of multiparticle production at the avalaible energies to an almost stochastic one  at the asymptotic energies. Transition to the reflective dynamics starts at the LHC energies and this fact makes searches of   its  signatures in multiparticle production, which were discussed above to be rather interesting .  As it was noted in \cite{intje} transition to reflective scattering mode would provide
a faster decrease of the energy spectrum reconstructed from extensive air showers in the cosmic rays measurements, i.e. it
  will result in  appearance of the knee in this spectrum. This knee   is  correlated with the slowing-down of the mean multiplicity growth discussed in \cite{slow} and should occur in 
  the same region of energy.

\end{document}